\documentclass[twoside,twocolumn]{article}

\usepackage{graphicx}

\usepackage{amsmath}

\usepackage[sc]{mathpazo} 
\usepackage[T1]{fontenc} 
\usepackage{microtype} 

\usepackage[english]{babel} 

\usepackage[hmarginratio=1:1,top=32mm,columnsep=20pt]{geometry} 
\usepackage[hang, small,labelfont=bf,up,textfont=it,up]{caption} 
\usepackage{booktabs} 

\usepackage{lettrine} 

\usepackage{enumitem} 
\setlist[itemize]{noitemsep} 

\usepackage{abstract} 

\usepackage{titlesec} 
\renewcommand\thesection{\Roman{section}} 
\renewcommand\thesubsection{\roman{subsection}} 
\titleformat{\section}[block]{\large\scshape\centering}{\thesection.}{1em}{} 
\titleformat{\subsection}[block]{\large}{\thesubsection.}{1em}{} 

\usepackage{fancyhdr} 
\usepackage{titling} 

\pretitle{\begin{center}\Huge\bfseries} 
\posttitle{\end{center}} 
\title{Radiation shielding composites using thermoplastic polymers mouldable at low temperature} 
\author{%
\textsc{John McMillan} \\[1ex] 
\normalsize Department of Physics and Astronomy, The University of
Sheffield\\ Sheffield, South Yorkshire, S3 7RH, Great Britain \\ 
\normalsize {j.e.mcmillan@sheffield.ac.uk} 
}
\date{\today} 
\renewcommand{\maketitlehookd}{%
\begin{abstract}
\noindent The formulation of a low temperature thermoplastic mouldable radiation shielding
composite is given.  The material is based on a low
temperature melting polymer filled with lead shot.  It can be easily moulded
to shape after immersing in hot water, or by using a hot air gun.
When cooled it is a rigid self supporting mass.
Alternative formulations to provide low background or low toxicity
gamma shielding or neutron
shielding are also considered.
\end{abstract}
}

\begin{document}

\maketitle

\section{Introduction}

Many medical and experimental procedures benefit from the
availability of radiation shielding that can be moulded
to shape without using metal cast at high temperature.
Formulations for shielding putties based on clays
and thermoset resins \cite{Fa92a,Ei94a,Sa04a,Ok05a} have been
published but these have the disadvantage that the material is not reusable.
A commercial product known as ``EnviroClay'' based on bismuth
held in some binding clay is available from Radiation Products Design
Inc.\ \cite{Ra}
but this remains soft and pliable at room temperature.

A thermoplastic shielding material was introduced by
Maruyama et~al.\ \cite{Ma69a}
which consisted of lead shot embedded in a matrix of a tooling wax
(M. Arg\"{u}eso \& Co., ``Rigidax'').  A similar composition,
 described by Haskell~\cite{Ha96a},
was a thermoplastic dental wax (Moyco Industries ``impression compound'')
loaded with bismuth powder, chosen for its low toxicity.
This last work suggests that ``a synthetic thermoplastic could be
substituted for the preferred hydrocarbon wax blend'' but,
presumably, at that time, no suitable material was available.

The current work describes the use of polycaprolactone as a binder
to produce novel thermoplastic shielding compounds for gamma, X-ray and
neutron applications.
The impetus was to improve shielding in low background
particle-astrophysics experiments.  Such experiments are typically
operated deep
underground within shielding castles constructed from lead blocks
and with an inner layer of ultra-pure copper blocks.   Inevitably, there
are ports and voids in the shields where pipework and cables pass
through the castles.  Further, the castles are customarily fabricated
from available or stock-sized blocks or ingots which rarely match the
experimental apparatus accurately.

There is a clear
requirement for a shielding material that will fill the spaces and
voids in the castle as precisely as possible.
Either cutting or casting metal to shape are possible but both are
problematic, especially in an underground laboratory.
Another approach is to use lead wool or shot held in bags or sacks,
but these have handling problems and the potential release of
finely divided material. Incorporating the shielding metal in a
polymer to provide a mouldable thermoplastic offers a simple
system of containment combined with flexibility of
use and re-usability.

In low background experiments, a key requirement is the absence of
radioactive impurities in the shielding,
consequently the formulations based on waxes
\cite{Ma69a,Ha96a} were not usable. These waxes contain high
proportions of mineral
fillers which would be expected to contain unacceptable levels of
uranium and thorium. EnviroClay \cite{Ra} was similarly rejected.

It is considered that the composites described here will have more
general applicability than the application for which they were
originally produced.

\section{Polycaprolactone}
Polycaprolactone is a rigid crystalline polymer, having the unit
formula (C$_{6}$H$_{10}$O$_{2}$)$_{n}$, which is remarkable
for its low melting point of 58--60$^\circ$C.  When molten it easily
wets uneven or greasy surfaces and is consequently much used as
a hot-melt adhesive.  It is soluble in a range of common solvents, is
intrinsically non-toxic and is bio-degradable.

A range of thermoplastic polycaprolactones are manufactured by
Ingevity \cite{In} as powders or pellets under the CAPA tradename.
These were previously sold under the Perstorp and Solvay brand names.
Various grades are available
with molecular weights from 10,000 to 80,000.
Small quantities of polycaprolactone (CAPA6800) in pellet form are
marketed for experimental and hobbyist
purposes under the tradename ``Polymorph'' in the
UK and this was the inspiration for the present study.  Similar
material is marketed
under the names ``Shapelock'', "Polydoh", "Friendly Plastic" and others.

Initial investigations of samples of Perstorp polycaprolactone
using a low-background Germanium detector indicate that its
radiopurity is high, comparable with other pure polymer materials.
Concentrations of uranium, thorium and potassium were all measured
as less than 30ppb by weight.

\section{Shielding for X-rays and gamma-rays}
To provide maximum shielding for X-rays and gamma-rays the
thermoplastic matrix should be loaded as highly as possible with
dense metal.  For the present project fine lead shot was chosen for
cheapness, but copper could be used to produce a low background
composite, while bismuth would produce a low toxicity
composite, but at higher cost.    Tungsten granules could also be
considered and this would offer a method of producing very
dense shields without the difficulty of machining tungsten metal.

The ratio of lead to binder was chosen to ensure that the
composite would have the maximum strength
available consistent with the maximum density, which
would be obtained when all voids between the lead are filled with polymer.
Assuming that the lead shot consisted of perfect identical spheres,
which were randomly stacked,
the highest packing fraction obtainable
is 0.64 \cite{Ja92a}.
The proportional mass of polymer required to bind the lead is given by:
\begin{equation*}
    \frac{m_{poly}}{m_{Pb}} =
    \frac{(1-0.64)\rho_{poly}}{0.64\rho_{Pb}}
     = 0.56\cdot\frac{1.1}{11.3}  = 0.054
\end{equation*}
Maruyama~et~al.\ \cite{Ma72a} point out that by using a mixture
of shot sizes, even higher packing fractions may be attained.
Specifically, with a mixture of two sizes having a ratio of diameters
of 0.22 and a mixing ratio of between 1:2--1.4 larger to smaller
pellets, a packing fraction of 0.72 was obtained.  More recent work
on packing fractions of particles with different distributions of
sizes was reviewed by Brouwers~\cite{Br06a}.
Such distributions were not
further investigated in the present study.

Lead based thermoplastic shielding composite was prepared using
Perstorp CAPA6506, a polycaprolactone with an approximate molecular
weight of 50000 produced as powder with a 600$\mu$m grain size.
It would be difficult to ensure a homogenous sample if pelleted polycaprolactone were used.

The lead shot was obtained from Calder Industrial Materials
Ltd.~\cite{Ca} and had an average size of
1.3mm, though it was clear that there was a distribution both in size
and shape.
The packing fraction of this material was measured as 0.63, only
marginally less than the theoretical prediction for uniform spheres~\cite{Ja92a}.

Initial 100g samples were prepared by weighing out the lead shot
and polycaprolactone and intimately mixing them by simply stirring.
They were then
heated in a shallow stainless steel vessel (known in the UK as a
``balti dish'') on a hot plate.
Production quantities of the composite, up to 1kg, were subsequently
made in larger stainless steel pans.
It was found advantageous to
include 50ml or so of water per 100g of lead to prevent the temperature
exceeding
100$^\circ$C, as the polycaprolactone decomposes at about 200$^\circ$C\
and smokes badly if overheated. The water also helped conduct the heat
uniformly to the composite mass and, to an extent, prevented the
composite from adhering to the pan.

A certain amount of stirring or kneading the material
was required to ensure that the
lead and polymer were adequately mixed.
At this stage the composite was a viscous toffee-like mass.
It could be removed from the
dish with a spoon or gloved hand and any water remaining on it would
run off.  The warm composite
was somewhat sticky and when
moulding the material, it was found that the mould should
either be greased or lined with polythene or mylar sheet.  On cooling, it was
found almost impossible to remove the composite from the mould if
this precaution was not taken.

When the composite is at the boiling point of water,
extreme care should be taken as getting the hot material
on bare skin would be likely to cause a painful scald.
However, when it had cooled to 60$^\circ$C\ it was found
that it could be easily worked to shape by hand and remained workable
for a few minutes.

\begin{figure}
\vspace{1mm}
\center
\includegraphics[width=\columnwidth]{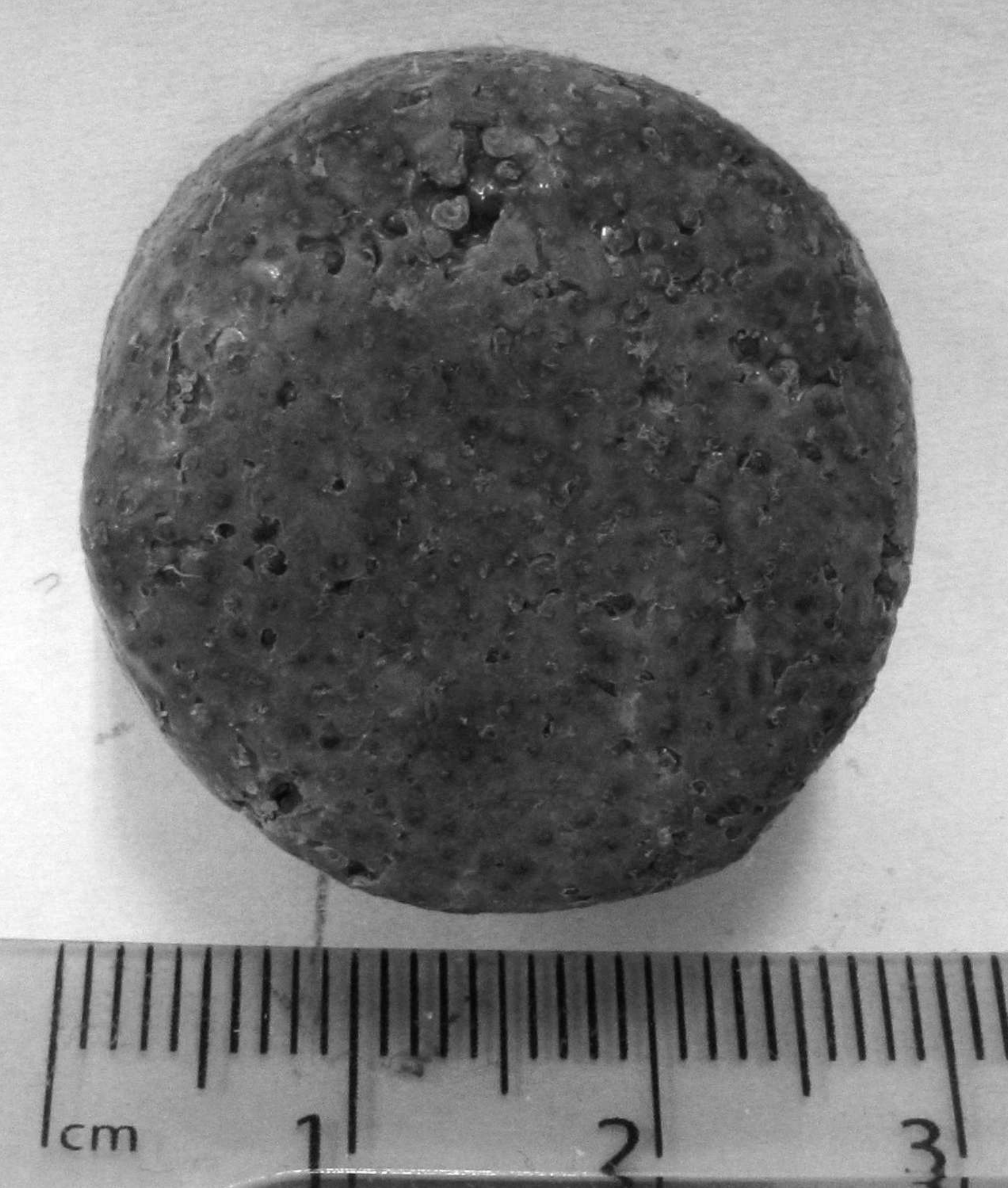}
\caption{Test sample of lead composite. }
\label{sample}
\vspace{-5mm}
\end{figure}

The cooled material was a tough resinous solid with
a measured density, $\rho_{comp} = 6.7$, which was slightly less than
the predicted density of 7.4.  Presumably a
small fraction of voids remain.
A small (66g) sample moulded into a round cornered
cylinder can be seen in figure~\ref{sample}.
At the surface, all
the lead particles are covered with a film of polymer which should assist in minimizing
contact with toxic metal.
Surprisingly, the composite was found to be electrically
non-conducting; each particle of lead being insulated from its
neighbours by films of polymer.

The composite could be remelted by simply returning it to the pan
and reheating.  Some experiments were made on moulding the composite
using a hot air gun.  This was possible but care had to be taken
since the composite had rather poor thermal conductivity
and it was found easy to overheat the material at the surface before the
core had melted.

Simple tests of the gamma shielding capability of the composite using
weak sources and a geiger counter
indicated that its properties were exactly that which would be
expected from reduced density lead.

\section{Shielding for neutrons}
Shielding of particle-astrophysics experiments against environmental neutrons
typically requires hydrocarbon material of at least
10cm thickness in order to ensure that the external neutrons are
thermalised and captured.  In the case of detector
systems larger than
a cubic metre, the requirement for neutron shielding material
is of order some tonnes, therefore there is the overriding
need for low cost material \cite{Mc05a}. As before, there is also
the requirement for high radiopurity of low background materials.
Similar considerations are also
applicable to other experimental disciplines where bulk neutron shielding
is required.

Sakurai \cite{Sa04a} and Okuno \cite{Ok05a} both describe the use of
thermoset resins loaded with either isotopically enriched lithium or
boron compounds to capture thermal neutrons.
For particle-astrophysics
applications, the difficulty of re-use of such material makes the use
of thermosets unattractive while the cost of large quantities of
these materials remains prohibitive.

Considerable use has been made of thick slabs of polypropylene or
high density polyethylene, which are widely available commercially but
remain relatively expensive.  A far cheaper option is to use raw
polyethylene or polypropylene pellets held in fabric bags or sacks or
contained within wooden shuttering \cite{Mc05a}.  The thickness of
material has to be increased to allow for the packing fraction
but this is rarely problematic.

The availability of a mouldable composite based on polymer pellets
would mean that the containment could be dispensed with.
A composite material can be produced using
polypropylene pellets embedded in water extended polyester (WEP) at
reasonable cost \cite{Mc05a}.
Unfortunately, this has the disadvantages mentioned
above with respect to thermosets combined with problems of
manufacture associated with the presence of styrene monomer in the
resin.

Pure polycaprolactone, being (C$_{6}$H$_{10}$O$_{2}$)$_{n}$
is an intrinsic neutron shielding material,
however, it is not a cost-effective choice, being approximately
seven times as expensive as polypropylene
pellets.  To reduce the cost,  a composite was produced using
polypropylene pellets bound together using polycaprolactone.
A composite in which all the voids were filled would have optimal
strength, while a composite with minimum binder would result in the
lowest cost.

The materials used were CAPA6506 polycaprolactone powder and Targor
Procom polypropylene pellets.  These latter were roughly cylindrical
with a diameter of about 4.5mm and an average of 3.6mm thick, though
there was considerable variation between pellets.
Experiments were performed to determine the optimum ratio of pellets
to binder.
The production method was similar to the lead version.
To produce large panels of this composite,
it should be possible to make a mould in which the ingredients are
heated by blowing steam through the material.

It was found that with low proportions of CAPA to
polypropylene beads, the resulting composite was not structurally
supporting and easily disintegrated.
Composite produced with 25g CAPA to 100g of
polypropylene beads or more was structurally sound, however, it
is of comparable price to commercially available thick
slabs of polypropylene.  While it is only marginally cost-effective,
it may have advantages in terms of easy
fabrication in complex shapes and in re-usability.

While this was not attempted, it is clear that quantities of boron or lithium
compounds, or minerals such as colemanite,
could be added to the composite without greatly affecting its mechanical
properties.  Other neutron absorbers such as compounds of gadolinium,
cadmium or europium could similarly be considered.  Polycaprolactone
could also be employed as an alternative to the epoxy resin binder
used in ``crispy mix'' boron-carbide based neutron shielding
composites \cite{Pu85a}, again giving a remouldable thermoplastic
material.

\section{Long term stability}
While no long term assessments of these composites have been
performed, a few general observations can be made.
Polycaprolactones are biodegradable, but this appears to be by
fungal attack in the presence of moisture.  If the material is
kept dry, it can be expected to last as long as any other plastic
material.
If polycaprolactone is exposed to very high doses (500kGy) of gamma
radiation the polymer will cross-link \cite{Da98a},
which would be expected to render the material stiffer and even
more viscous in the molten state and would presumably lead to
embrittlement in the solid phase.

\section{Acknowledgements}
The author would like to acknowledge the assistance of Solvay in
providing samples of polycaprolactone.

\bibliographystyle{unsrt}
\bibliography{radlute_arxiv}
\end{document}